\documentclass[doublecol]{epl2} 
\usepackage{epsfig} 

\def\be{\begin{equation}}
\def\ee{\end{equation}}
\def\bc{\begin{center}} 
\def\ec{\end{center}}
\def\bea{\begin{eqnarray}}
\def\eea{\end{eqnarray}}

\title{Aligning graphs and finding substructures by a cavity approach}
\author{S. Bradde\inst{1,2} \and A. Braunstein\inst{2} \and H. Mahmoudi\inst{3}  \and F. Tria\inst{3} \and M. Weigt\inst{3} \and R. Zecchina\inst{2}}
\shortauthor{S. Bradde \etal}
\institute{                    
  \inst{1} SISSA - via Beirut 2/4, I-34014 Trieste, Italy and INFN Sezione di Trieste, Italy \\
  \inst{2} Politecnico di Torino - Corso Duca degli Abruzzi 24, I-10129 Torino, Italy\\
  \inst{3}  Institute for Scientific Interchange - Viale Settimio Severo 65, I-10133 Torino, Italy
}
 {Spin-glass and other random models}

\abstract{
We introduce a new distributed algorithm for aligning graphs or
finding substructures within a given graph. 
It is based on the cavity method and is used to study the maximum-clique and the graph-alignment
problems in random graphs. 
The algorithm allows to  analyze large graphs and may find applications in fields such as computational
biology.  As a proof of concept we use our algorithm to align the
similarity graphs of two interacting protein families involved in
bacterial signal transduction, and to predict actually interacting
protein partners between these families.}

\begin{document}

\maketitle

Over the last decade, the use of graphs for the description of 
relations between components of complex systems has become
increasingly popular \cite{networks}. 
However, most part of the current literature concentrates on (computationally accessible)
local characteristics like node degrees,  whereas the full exploitation of more
global properties of large networks remains frequently elusive
due to their  inherent algorithmic complexity.  
Most often studying global properties requires solving NP-hard problems or even harder 
problems if some form of uncertainty or  lack of information  is included in the definition 
of the problem.  In  both cases  heuristic algorithms need to be developed. 
Specific examples, covered by this article, include the comparison of two
different networks, the so-called {\it graph-alignment problem} (GA)
\cite{Berg,Li,Singh}, and the {\it sub-graph isomorphism} (SGI), as a 
particular case of which we consider the widely studied maximum-clique 
problem \cite{karp1972,boppana,Bollobas}.

Recently, there has been a lot of interest in distributed
algorithms to deal with optimization problems over networks. In the
context of statistical physics a new generation of algorithms has been
developed (e.g. \cite{MePaZe,BrMeZe2005}) that have shown promising 
performance on several
applications (for a review, see \cite{marc}). These techniques are based on the so called cavity
method and are known as message-passing (MP) algorithms. 
They are fully distributed and easy to run on parallel machines.
A recent result in this framework is an algorithm for finding a connected
sub-graph of a given graph which optimizes a given factorized cost
function \cite{steiner}.

Here we aim at making a step forward by introducing new techniques for SGI 
and GA. We develop two alternative MP strategies and test their
performance on three sample problems. The first two are well-defined theoretical
benchmarks, where our results can be compared to rigorous bounds: (i) the 
maximum-clique problem in random graphs for the SGI problem; and (ii) the alignment 
of two random graphs of controlled similarity. The third sample problem is thought 
as a proof-of-concept application in computational biology: We study (iii) the
alignment of the similarity networks of two interacting protein-domain
families involved in bacterial signal transduction, to identify actual 
signaling pathways. This case, involving large networks of $>$2500 nodes, exploits 
co-evolutionary processes between interacting proteins to identify interaction
partners \cite{Ramani}.

{\it The model} --- Both problems, SGI and GA, can be put into the
common framework of matching two graphs of possibly different
size. Let $G=(V,E,w)$ and $G'=(V',E',w')$ be two weighted graphs with
nodes $V, V'$, edges $E, E'$ and edge weights $w, w'$. In the applications
shown in this letter, weights are non-negative, but this is not a 
necessary condition for the applicability of the message-passing algorithms.
In the case of unweighted graphs, we assume $w$ and $w'$ to describe the 
adjacency matrices, i.e. weights are one if an edge is present between two 
vertices, and zero else. Further more we denote the node number by $N=|V|$
($N'=|V'|$), and the edge number by $M=|E|$ ($M'=|E'|$). Neighbors of
a node $i$ are assembled in $\partial i$. To facilitate notation,
primed quantities (in particular node indices) will always refer to
$G'$. Without loss of generality, we assume $N\le N'$.

The problem is now to find an {\it injective mapping} $\pi: V\to V'$
between the nodes of $G$ and $G'$. This mapping minimize the cost function 
(Hamiltonian) 
\be
\label{eq:H}
{\cal H}(\pi) = -\sum_{(i,j)\in E} w_{ij} w'_{\pi_i\pi_j} -\sum_i
c_{i\pi_i} 
\ee 
Note that in the case of unweighted graphs the first
term counts the number of overlapping links in the graphs $G$ and
$G'$.  The $c_{ii'}$ denote similarities between nodes of the two
graphs, i.e.  they provide a local bias for the mapping $\pi$ which,
in physical terms, represents a local external field. Computationally
this problem is very hard: The complexity of a simple enumeration is
${\cal O}(N'^N)$, i.e. it is growing more than exponentially for
growing $N$ and $N'$. It is a special case of the so-called {\it
quadratic assignment} problem, which would contain any cost
$e(i,j,\pi_i,\pi_j)$ in the first sum of the Hamiltonian. The
generalization of our algorithm to this case is straight-forward.
A major problem is to implement the injectivity constraint: for
$i\neq j$ also $\pi_i\neq\pi_j$ has to hold. We introduce two 
strategies to treat this constraint within approximate algorithmic
approaches based on message passing:

{\it (i)} We directly introduce this constraint for each pair of nodes
in $G$, using a complete graph of these vertices, where each
link carries the constraint. The Boltzmann distribution
(at finite formal temperature $1/\beta$) reads 
\be
\label{eq:P1}
P^{(1)}(\pi) = e^{-\beta {\cal H}(\pi) } \prod_{i,j\in V; i<j} (1-\delta_{\pi_i,\pi_j})
\ee
The application of message passing to this formulation of the
problem requires the exchange of vectorial messages of dimension $N'$
along all links of the complete graph of constraints, so the complexity
becomes ${\cal O}(N^2N')$, cf.~Eq.~(\ref{eq:bp1a}-\ref{eq:bp1b}) below. 

{\it (ii)} We relax the constraint, i.e. we consider arbitrary mappings 
$\pi:V\to V'$ and then we couple a chemical potential $p$ to the 
number $N_\pi = |\pi(V)|$ of images of $\pi$:
\be
\label{eq:P2}
P^{(2)}(\pi) = e^{-\beta {\cal H}(\pi) + \beta p N_\pi } \ee 
To be usable in a message-passing approach, we further express the
image number as $N_\pi=\sum_{i'\in V'} \chi_{i'}(\pi)$, with 
$\chi_{i'}(\pi)=1$ if there exists an $i\in V$ having $i'=\pi_i$ as it's 
image, and $\chi_{i'}(\pi)=0$ else. For
sufficiently large but finite values of $p$, the ground states have
$N_\pi=N$, and injectivity is restored. This leads to a slightly more
involved message-passing algorithm (cf. below) whose time complexity goes 
down to ${\cal O}((N+M)N')$; this formulation is favorable in particular
for large sparse graphs $G$.

{\it Message-passing algorithms} --- An exact treatment of the two
approaches gives equivalent results for $\beta\to\infty$, but it is
infeasible for large $N$ due to the super-exponential time
complexity. Here we develop two heuristic algorithms using MP.

{\it (i)} The first algorithm is a straight forward application of
belief propagation (BP) to the problem defined in
Eq.~(\ref{eq:P1}). Messages are exchanged between any two nodes in
$V$, and they are of dimension $N'$: \bea
\label{eq:bp1a}
P_{i\to j} (\pi_i) &\propto& e^{\beta c_{i\pi_i}} \prod_{k\neq i, j} Q_{k\to i}(\pi_i) \\
\label{eq:bp1b}
Q_{i\to j} (\pi_j) &\propto& \sum_{\pi_i} (1-\delta_{\pi_i,\pi_j})
e^{\beta w_{ij} w'_{\pi_i\pi_j}} P_{i\to j} (\pi_i)    \nonumber
\eea
Messages are standard cavity probabilities and biases \cite{marc}, with 
$P_{i\to j}(\pi_i)$ being the marginal probability of node $i$ in the cavity
graph which is constructed by removing $j$ from the node set $V$ (the cavity
graph is thus a complete graph of $N-1$ nodes). Message $Q_{i\to j}(\pi_j)$
describes the bias induced by node $i$ on the mapping of node $j$, including both 
the Boltzmann factor of the weights of aligned edges $(i,j)$ and $(\pi_i,\pi_j)$
and the injectivity constraint $(1-\delta_{\pi_i,\pi_j})$. The BP 
equations can be solved iteratively, and the marginal probability 
that node $i\in V$ chooses $\pi_i\in V'$ as its image is given by
\be
\label{eq:marg1}
P^{(1)}_i(\pi_i) \propto e^{\beta c_{i\pi_i}} \prod_{k\neq i} Q_{k\to i}(\pi_i)\ .
\ee

{\it (ii)} Giving a full derivation of the BP equations for analysing
Eq.~(\ref{eq:P2}) goes beyond the scope (and space limitations) of this letter.
We will give, however, some indications about the main steps: The factor graph
corresponding to $P^{(2)}(\pi)$ has $N$ variable nodes $i\in V$, each one carrying
a $N'$-state spin variable $\pi_i$. There are two types of factor nodes: The
first type corresponds to the $M$ edges $(i,j)\in E$ of graph $G$ and measures the
alignment weight; the second type corresponds to each of the $N'$ possible image 
nodes $i'\in V'$ and depends on the indicator $\chi_{i'}(\pi)$ if node $i'$ is 
selected as an image or not. These factor nodes are {\it a priori} problematic, 
since each one is connected to all
variable nodes $i\in V$. Applying na\"ively BP requires {\it a priori} a summation 
over ${\cal O}(N'^N)$ terms. However, the symmetry structure of the problem is 
the same as the one of (soft-constraint) affinity propagation 
\cite{Dueck,LeoneSumedhaWeigt,Leone2}: Even if messages from a factor node $i'$ 
to a variable node $i$ are given as $N'$ dimensional vectors of the form
$\tilde A_{i'\to i}(\pi_i)$, they contain just two different entries for $\pi_i=i'$
and $\pi_i\neq i'$. The before-mentioned sums can be performed analytically, cf.
\cite{Dueck,LeoneSumedhaWeigt,Leone2}. Here we state only the final equations: 

\bea
\label{eq:bp2}
&&A_{i'\to i} = \left[ 1-(1-e^{-\beta p}) \prod_{j\neq i} (1-B_{j\to i'})  \right]^{-1}
\nonumber\\
&&B_{i\to i'} =\! \!  \left[ 1\! \!  +\! \sum_{j' \neq i'} e^{\beta (c_{ij'}-c_{ii'})} A_{j'\to i} 
\prod_{j\in \partial i} \frac{Q_{j\to i}(j')}{Q_{j\to i}(i')} \right]^{-1}
\nonumber\\
&&P_{i\to j}(\pi_i)\!  \propto \!\! \! \left[ 1\! \! +\! \! \! \sum_{i'\neq\pi_i}\! \!  
e^{\beta (c_{ii'}-c_{i\pi_i})} \! 
\frac {A_{i'\to i}}{A_{\pi_i\to i}} \! \! 
\prod_{k\in \partial i\setminus j} \! \! \frac{Q_{k\to i}(i')}{Q_{k\to i}(\pi_i)} \right]^{-1}
\nonumber\\
&&Q_{i\to j} (\pi_j)\! \!  \propto \sum_{\pi_i} 
e^{\beta w_{ij} w'_{\pi_i\pi_j}} P_{i\to j} (\pi_i) 
\eea
Messages $A$ and $B$ are exchanged between the nodes of the two graphs, they have a nice
intuitive interpretation: $B_{i\to i'}$ is a {\it request} of $i$ to $i'$; it measures in 
how far $i$ would like to select $i'$ as his image. According to the second of 
Eqs.~(\ref{eq:bp2})it depends on the node 
similarity $c_{i,i'}$ as compared to the similarities $c_{i,j'}$ of $i$ to all other 
$j'\neq i'$, and on the $Q$-messages containing the alignment weight for links $(i,j)\in G$. 
Message $A_{i'\to i}$ indicates the {\it availability} of $i'$ to become image of $i$. 
It is large if the requests $B_{j\to i'}$ from other nodes $j\neq i$ to $i'$ are small,
favoring therefore large $N_\pi$.
Messages $Q$ and $P$ are exchanged along the edges of $G$. After finding a fixed point 
of these equations, the marginal for $i\in V$ is given by
\be
P^{(2)}_i(\pi_i) \propto e^{\beta c_{i\pi_i}} A_{\pi_i\to i} (1-A_{\pi_i\to i})^{N-1}
\prod_{j\in\partial i} Q_{j\to i}(\pi_i) \ .
\ee

Being of approximate nature, it is not clear that these two MP strategies lead to the same 
results applied to the same graphs, but in practice we did not observe systematic differences.
The major difference is the time complexity: Whereas for relatively small $G$ (as in the
sub-graph isomorphism discussed below) the first strategy was found to be faster, the two
applications concerning larger but sparse $G$ (alignment of random graphs and 
protein-similarity graphs) are solved faster by the second strategy.

The BP equations are fixed-point equations, and can be solved by
iterating Eqs.~(\ref{eq:bp1a})-(\ref{eq:bp1b}) resp.~(\ref{eq:bp2}). This is not
guaranteed to converge, nor there is a guarantee that only one fixed
point exists. Since in optimization we are interested in
constructing one solution, this problem can be circumvented by 
enforcing convergence through the so called soft decimation or 
reinforcement technique \cite{BrZe2006b}. 

In the case of the first algorithm (Eqs.~\ref{eq:bp1a}-\ref{eq:marg1}) it 
amounts to multiplying  
at each step the right hand side of Eqs.~(\ref{eq:bp1a}) and (\ref{eq:marg1}) by the term
$\left[P_i^{(1)}(\pi_i)\right]^{\gamma_t}$, i.e. the power of the marginal probability, 
as computed in the previous step with the (modified) Eq.~(\ref{eq:marg1}),
to a scalar time dependent $\gamma_t$. For the simulations in  this work, we used $\gamma_t = 1 - \alpha^t$ for $\alpha$ close to one.

Note that BP bears some similarity with the IsoRank algorithm \cite{Singh}. There are,
however, some crucial differences: (i) BP is derived from a variational minimization
of cost function (\ref{eq:H}); (ii) IsoRank is a mean-field algorithm, whereas BP
is based on the more precise Bethe approximation; (iii) IsoRank has no explicit
control over the injectivity of the resulting alignment.

{\it Finding maximum cliques} --- In the following, we are going to test BP 
on three sample problems. The setting of the parameters is $\beta=\infty$ for version (i) and $\beta=10$ and $p=-40,-100$  for version (ii) of the algorithm.

The first application concerns finding the maximum clique in a given graph. 
On the computational side, this is indeed a root problem
being both NP-complete \cite{karp1972} and difficult to approximate \cite{boppana}. 
Within our previous notation, graph $G$ now is a complete graph with $M=N(N-1)/2$, 
and we are trying to embed it into a second graph $G'$ of normally much larger order $N'$.
Node similarities $c_{i,i'}$ are set to zero.
In the specific case that $G'$ is a random graph with $N'$ nodes and edge probability
$N'^{-\alpha}$, $\alpha\in(0,1)$, rigorous bounds for the maximum clique size $cl(G')$ 
are known \cite{Bollobas}: 
\be
\left[ k_0 - 2 \frac{\log \log N'}{\log N'} \right] \leq cl(G') \leq
\left[ k_0 + 2 \frac{\log \log N'}{\log N'} \right]
\ee
with
\be\label{ko}
k_0(N') = \frac 2\alpha + 2\frac{\log \alpha}{\alpha\log N'} + 
2\frac{\log \frac e2}{\alpha\log N'} + 1 + o(1)\ .
\ee
As usual in random-graph theory, these bounds hold with probability tending to one 
in the limit $N'\to\infty$ of large target graphs $G'$.
Eq.~\ref{ko} is derived from the expected number of cliques of a given size, which
has to be of order 1. This expected number allows us also to estimate the 
number of smaller cliques in $G'$, which can be directly compared to the Bethe entropy 
calculated by BP.

\begin{figure}
\epsfig{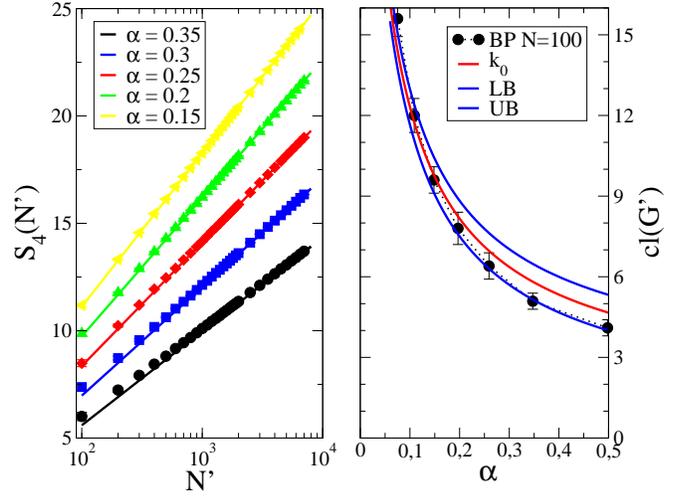}
\caption{In the left panel, we report the entropy of cliques of size 4, as a function of
the order $N'$ for various values of $\alpha$. Symbols are BP results
(each symbol averaged over 10 random graphs),
full lines give the logarithm of the expected number of such cliques. On the right, we
display the largest clique found by BP in graphs of size $N'=100$, as a function of $\alpha$
(each symbol averaged over 10 random graphs). Results are consistent with theoretical bounds.}
\label{fig:SGI}
\end{figure}
Results of this comparison are presented in Fig.~\ref{fig:SGI}.
 Finite size effects set in for small entropy values,
which correspond to the logarithm of the maximum cliques number,  
while for larger  $N'$ the BP results perfectly coincide with the theoretical predictions.
We note that in the case of cliques, and more in general for highly symmetric sub-graphs,  it is possible to exploit symmetries to simplify the BP equations. The resulting equations have slightly different convergence properties compared to the generic BP ones.
The largest cliques size found by our algorithm are in agreement with the (fairly tight) 
rigorous bounds also for small value of $N'=100$, see right panel of Fig.~\ref{fig:SGI}.
In the explicit constructions of cliques  both the decimation and the reinforcement 
techniques have been used (the latter one being substantially faster).

Finding and couting small subgraphs is one of the major steps in the search for
network motifs \cite{Milo,Berg2,Picard}. In contrast to exhaustive algorithms 
\cite{Milo,Picard}, BP 
is able to handle the problem even for medium size subgraphs. Further more, working 
at finite temperature allows for non-perfect alignments, an thus for identifying, e.g., 
dense subgraphs instead of perfect cliques. A detailed exploration of this possibility 
goes beyond the scope of this letter.

{\it Aligning sparse random graphs} --- In the second problem, we use
BP to align two sparse random graphs $G$ and $G'$ with identical
numbers of nodes, $N=N'$, and links, $M=M'$. The injective mapping
$\pi$ thus becomes a permutation. To study the best GA as a function
of the inherent similarity between the two graphs, we construct $G$
and $G'$ such that they have $M-M_{rand}$ links in common, the other
$M_{rand}$ are chosen independently in the two graphs \cite{Kolar}.
Note that for $M_{rand}=0$, the problem reduces to identify an
isomorphism between the graphs. Due to the specific 
construction this isomorphism is trivially given by the identical
permutation, $\pi_i=i$, but this information is neither known nor
exploitable by the algorithm; it serves only for a simple evaluation
of the simulation results.
For $M_{rand}=M$, the two graphs are independent, and the number of
alignable links is monotonously decreasing with $M_{rand}$.

\begin{figure}[htb]
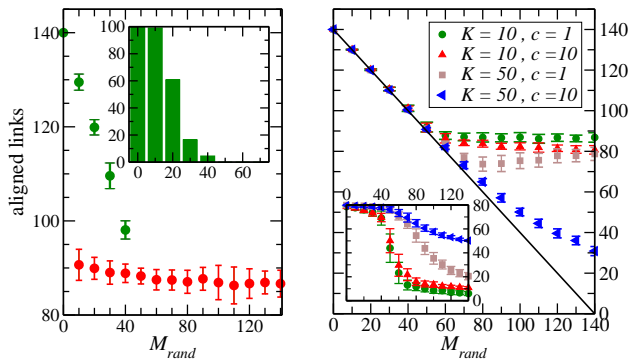

\vspace{5mm}
\includegraphics[width=0.42\columnwidth]{fig2a.eps}
\hspace{3mm}
\includegraphics[width=0.455\columnwidth]{fig2b.eps}
\caption{The number of aligned links as a function of the number
$M_{rand}$ of independent links in $G$ and $G'$, for $N=N'=80$ and
$M=M'=140$ (each data point averaged over 30 random-graph pairs and 5
BP runs). First panel: Results for alignment without node similarities
($K=0$). For positive but not two large $M_{rand}$ the solutions fall
into two classes: Green symbols represent close-to-identical permutations
of high number of aligned links, red symbols represent results which
are close to the alignment of two independent graphs. The inset shows 
the fraction of BP runs leading to a
close-to-identical permutation. Second panel: Results for different
values of the similarity parameters $K$ and $c$, the best performance
is obtained for $K=10$ and $c=1$ (small bias). The inset shows the
number of self-aligned nodes, it drops when the bias is not too large
and $M_{rand}>\bar M_{rand}$.}
\label{fig:GA}
\end{figure}

Results for the alignment of $G$ and $G'$ without node similarities
($c_{i,\pi_i}\equiv 0$) are given in the first panel of
Fig.~\ref{fig:GA}. For $M_{rand}=0$, BP always identifies correctly the
isomorphism between the two graphs. In the other limiting case,
$M_{rand}=M$ the number of aligned links is much smaller, and depends
on the graph realization and the initial condition of the BP messages.
In between the two extremes we find a transition in the algorithmic
behavior at some $\bar{M}_{rand}$ where the identical permutation has
the same value of $\cal H$ as the best alignment of two independent
graphs. Above $\bar{M}_{rand}$, the number of aligned links is found
to be almost constant, and equals the independent-graph case. For $0 <
M_{rand} < \bar{M}_{rand}$ the BP solutions fall into two different
classes: one being close to the identical permutation with a high
number of aligned links (green symbols in Fig.~\ref{fig:GA}), and one 
having $\cal H$-values coherent with the alignment of two independent 
graphs (red symbols). The relative fraction of the first case is 
shown in the inset, it decreases when approaching $\bar M_{rand}$.

The behavior changes, when node similarities (cf.~Eq.~(\ref{eq:H}))
are used to bias BP toward the identical permutation. We set
$c_{i,\pi_i}=c \delta_{i,\pi_i}$ for $i=1,...,K$, and $c_{i,\pi_i}=0$
for $i>K$, with $K\in\{0,...,N\}$. The parameter $K$ controls the
number of biased nodes, and $c$ controls the strength of this bias.
Interestingly, already a low value of $K$ is sufficient to let BP
always find the lower energy solutions in the region $0<M_{rand} <
\bar{M}_{rand}$.  Results for different values of $K$ and $c$ are
shown in the second panel of Fig~\ref{fig:GA}. We observe that
excessively high values of $K$ or $c$ decrease the performance of the
algorithm for large $M_{rand}$ by forcing it towards solutions of 
less aligned links but more self-aligned nodes.

{\it Finding interacting protein partners from multi-species sequence data} ---
Despite the fundamental importance of protein-protein interactions in most
biological processes, identifying interaction partners is
experimentally and computationally a major problem. As a proof-of-concept 
application for GA we consider two
signaling proteins, namely a {\it histidine sensor kinase} (SK) and a {\it 
response regulator} (RR). Their interaction forms the central part of 
two-component signal transduction (TCS), which is the most prominent signal
transduction mechanism in bacteria. Each bacterium contains ${\cal O}(10)$
interacting SK/RR pairs forming different TCS pathways; and the
necessity to trigger the correct answer for each specific extracellular signal
forbids crosstalk between pathways. So, even if all different SK
in one species are homologous (and therefore structurally and functionally
similar), and the same is true for all RR, only specific samples of these
two {\it protein families} interact. Our question here is, if GA can help
to identify interaction partners.

We start from a large collection of SK 
sequences extracted from hundreds of bacterial genomes, and a second large 
collection of RR sequences coming from the same bacteria, and we aim at
extracting interacting SK/RR pairs, exploiting sequence similarities 
of proteins inside each family. The basic idea is simple: Two SK with very similar
amino-acid sequences will (due to their probably recent common evolutionary 
origin) interact with two similar RR. Globally spoken, an 
alignment of two similarity networks - one for the SK family, one for the 
RR family - might be able to pair a large fraction of all those SK and RR 
which actually belong to common TCS pathways \cite{Ramani}. Our data set 
consists of two multiple-sequence alignments (with gaps) for 2546 SK and 
2546 RR proteins from 231 genomes \cite{Weigt}. They are selected such that, 
due to the frequent coding of an entire TCS in one operon, the correct 
mapping is known, and can be used {\it a posteriori} to verify our GA results.

Similarity networks for each protein family are constructed as
$k$NN graphs: Each protein is linked to the $k$ most similar
proteins, where similarity is measured via the Hamming distance
$d_{ij}$ between the aligned aminoacid sequences of two proteins $i$ and $j$. 
The link weight is given as
$w_{ij} = \exp{[-{d^2_{ij}}/{d^2_k}]}$, with $d_k$ being the average
distance between each protein and its $k$th neighbor. One might use more
sophisticated distance measures (e.g. alignment scores), but due to the
proof-of-concept character of this application we have chosen the simplest
possible measure. To identify interaction partners, we
must align only proteins inside the same species (formally implemented by
$c_{i,i'}=-\infty$ for all $i$ and $i'$ belonging to different
species). Finally, we have introduced various amounts of
information about real interaction partners, by randomly
introducing positive similarities between a number of actual
interaction partners (training set). The results are summarized in
the following table for different $k$ and training-set sizes.
Error bars result from an average over different random training sets. 
The values display the fraction of correctly aligned protein pairs in 
between all proteins not being in the training set.

\begin{table}[htb]
\begin{tabular}{|c|c|c|c|}
\hline
training set  & 3NN, $k=3$ & 6NN, $k=6$ & 9NN, $k=9$ \\
\hline
2000 & 88.7 $\pm$ 1.7 & 89.8 $\pm$ 1.9 & 90.5 $\pm$ 1.7\\
1000 & 76.2 $\pm$ 1.3 & 78.7 $\pm$ 1.0 & 79.6 $\pm$ 0.8 \\
500 & 67.4 $\pm$ 1.9 & 73.1 $\pm$ 1.4 & 75.0 $\pm$ 1.0 \\
0 & 48.1 & 58.9 & 64.7\\
\hline
\end{tabular}
\label{table1}\end{table}

We note that even without training set, almost 65\% of all
proteins are correctly matched (for $k=9$). This number has to be
compared to a random matching, where only $231/2546 \sim 9\%$
correct matchings would be expected. The introduction of a training set
improves strongly the performance, for a training set of 2000 protein
pairs, about 90\% of the remaining 546 proteins are correctly
aligned. These results beautifully demonstrate that the original
idea to exploit sequence similarity of proteins across species 
is actually providing information about who is interacting with whom.

{\it Conclusion} --- In this letter we have presented a distributed
(parallel) algorithm for graph-alignment problems.  The new
technique is based on the cavity method and allows to deal efficiently
with optimization problems under (global) topological constraints.
The results on famous problems such as sub-graph isomorphism and graph
alignment on random graphs are in remarkable agreement with known
rigorous bounds.  
Problems of this type are often encountered in the
analysis of large-scale data in many fields of science, computational biology in  first place.

\bibliographystyle{unsrt}
\bibliography{biblio}
\end{document}